\documentclass[9pt,twocolumn,twoside]{opticajnl}
\journal{opticajournal} % use for journal or Optica Open submissions
\setboolean{shortarticle}{false}
\usepackage{lineno}
\usepackage{nicefrac}

\title{Microchip semiconductor membrane external-cavity surface-emitting laser}
\author[1,2,*]{Jakob~Hirlinger-Alexander}
\author[2]{Michael~Scharwaechter}
\author[1]{Franzisca~Bader}
\author[1]{Julius~Steck}
\author[2]{Matthias~Seibold}
\author[2]{Marco~Werner}
\author[2]{Roman~Bek}
\author[3]{Hermann~Kahle}

\affil[1]{Institute for Functional Nanosystems, Ulm University, Albert-Einstein-Allee~45, 89081~Ulm, Germany}
\affil[2]{Twenty-One Semiconductors GmbH, Kiefernweg~4, 72654~Neckartenzlingen, Germany}
\affil[3]{Department of Physics \& Astronomy, The University of New Mexico, 210~Yale~Blvd~NE, 87106~Albuquerque, New~Mexico, USA}

\affil[*]{jakob.hirlinger-alexander@uni-ulm.de}

\begin{abstract}
We demonstrate the first microchip semiconductor membrane external-cavity surface-emitting laser. This compact type of laser consists solely of a semiconductor gain region present as a micron-thin membrane, sandwiched between two transparent heat spreaders. The heat spreaders have a highly reflective coating on their outer facets, which assembles the laser's plane-parallel solid-state cavity with a total length of just $\sim$\,1\,mm. One of the coatings with slightly reduced reflectivity acts as outcoupling mirror. The microchip membrane external-cavity surface-emitting laser (microchip MECSEL) is optically pumped with a standard fiber-coupled diode laser module emitting at 808\,nm and stabilizes itself due to an occurring thermal lens. More than one watt of continuous wave output power around 1123\,nm and a record value in fitted slope efficiency of $\sim$\,51.4\,\% with MECSELs, while maintaining excellent beam quality (TEM$_{00}$, $M^{2}$\,<\,1.05), is demonstrated. Important properties of semiconductor lasers such as the efficiency, beam quality, and polarization were investigated. Further, this setup was used to characterize the thermal lens and it's dependence on the absorbed pump power in the microchip MECSEL. Such systems represent an attractive solution, when high-power output at customizable emission wavelength with excellent beam quality is needed in combination with very compact built size.

\end{abstract}

\setboolean{displaycopyright}{false} % Do not include copyright or licensing information in submission.

\begin{document}

\maketitle

\section{Introduction}
Semiconductor membrane external-cavity surface-emitting la\-sers (MECSELs) were realized about a decade ago for the first time \cite{Yang.Albrecht.ea_2015,Kahle.Mateo.ea_2016} to improve the thermal impedance of the already well established vertical-external-cavity surface-emitting lasers (VECSELs) \cite{Guina.Rantamaeki.ea_2017, Jetter.Michler_2021}.
First, a semiconductor gain region as known from VECSELs, but no monolithically integrated distributed Bragg reflector (DBR), was bonded to a single transparent heat spreader and used as a gain element in an external laser cavity \cite{Yang.Albrecht.ea_2015}. Shortly after, the double side heat spreader approach was realized \cite{Kahle.Mateo.ea_2016,Broda.Kuzmicz.ea_2017}, which was found to be thermally superior \cite{Yang.Albrecht.ea_2015, Phung.TatarMathes.ea_2022} over single side configurations \cite{Yang.Albrecht.ea_2015, Yang.Albrecht.ea_2016, Mirkhanov.Quarterman.ea_2017}.
But MECSELs not only improve the thermal management of a semiconductor gain region present as a membrane of $\sim$\,1\,\textmu m in thickness, which results in very low thermal resistances \cite{Broda.Kuzmicz.ea_2017, Kahle.Penttinen.ea_2019, Phung.TatarMathes.ea_2022, Schuchter.Huwyler.ea_2024}. The fact that a monolithically integrated DBR is obsolete in this new laser configuration, allows now to realize vertically emitting semiconductor lasers at wavelengths, which were not possible to realize before. For example, this is the case due to the lack of sufficient refractive index contrast in the InP material system. Omitting the DBR and realizing optically pumped InP based lasers in a membrane configuration finally allowed to reach the 1.6 to 1.8\,\textmu m \cite{Broda.Jezewski.ea_2020,Broda.Jezewski.ea_2021} wavelength region. Additionally, the lack of a monolithically integrated DBR has further implications.
%\cite{Yang.Albrecht.ea_2016}
It means that the phase of the optical field is no longer locked within the gain region due to close proximity of a DBR, and the laser has more freedom to establish a good overlap between the standing wave optical field and the quantum wells (QWs) or quantum dot layers \cite{Phung.TatarMathes.ea_2021} within the gain membrane. This typically results in wider tuning ranges, allowing the laser to fully benefit of the naturally broad gain of semiconductors \cite{Yang.Albrecht.ea_2016}. Furthermore, it allows the realization of new heterostructure designs \cite{TatarMathes.Phung.ea_2023}, where the impact of resonance of the gain region's sub cavity on the laser's performance is reduced. This in turn increases the error tolerance in active region design, which stands in full contrast to the well known VCSELs\cite{Raja.Brueck.ea_1989}. Another novel heterostructure design \cite{Rajala.TatarMathes.ea_2024} allows multiple types of QWs in the same active region creating the possibility of continuous wave broadband tuning. In terms of power scaling, the MECSEL approach also enables double-sided \cite{Kahle.Penttinen.ea_2019,Rajala.TatarMathes.ea_2024} or multi-pass \cite{Priante.Zhang.ea_2022} pumping, which results in better charge carrier distribution within the gain membrane and allows thicker gain membranes with more QWs to be adequately pumped.\\
When otherwise device compactness becomes important and highest output powers \cite{Priante.Zhang.ea_2021,Priante.Zhang.ea_2022} are not necessarily needed, the benefits of an open external cavity can be sacrificed in favor of a monolithic cavity to suppress the influence of environmental noise \cite{Lee.Moriya.ea_2023} reaching record narrow linewidth values \cite{Moriya.Lee.ea_2024}. If one wants to further shrink the footprint of the monolithic cavity, a microchip configuration of a MECSEL is possible as well. Microchip lasers were first realized with classical solid-state gain material \cite{Zayhowski.Mooradian_1989} and soon reached significant output power values \cite{Nabors.Sanchez.ea_1992}. Later, a semiconductor variant called microchip VECSEL \cite{Hastie.Hopkins.ea_2003b,Smith.Hopkins.ea_2004,Hastie.Morton.ea_2005,Kemp.Maclean.ea_2006,Park.Jeon_2006,Laurand.Lee.ea_2007} was realized. The highest output power of such a microchip VECSEL was 1\,W at an emission wavelength of $\sim$\,1\,\textmu m \cite{Laurand.Lee.ea_2009}.\\
In this work we present the first microchip MECSEL, where the solid-state cavity consists of two plane parallel silicon carbide (SiC) heat spreader platelets with the semiconductor gain membrane sandwiched in between. In this paper a detailed characterization of this novel device is presented.

\section{Sample and experimental setup}
\label{sec:sampex}
\subsection{Gain structure \& microchip assembly}
%\textcolor{red}{DAS MACHTE DER ROMAN: 
The multi-quantum well gain structure was fabricated by metal-organic vapor-phase epitaxy in a 4-inch shower-head reactor. It contains InGaAs QWs embedded in GaAs spacer layers and GaAsP layers for strain compensation. The gain membrane is enclosed by GaInP window layers to prevent surface recombination, and an AlGaAs layer between the gain membrane and the (100)-GaAs substrate facilitates the substrate removal process. Overall, the gain structure is similar to the one described in Ref.\,\cite{Priante.Zhang.ea_2021}.\\
After growth, the wafer was fusion-bonded \cite{Cole.Zhang.ea_2013} to a nominally 500\,\textmu m thick SiC heat spreader, the substrate was removed in a wet-chemical etching process and another 500\,\textmu m SiC heat spreader was fusion-bonded to the second side of the $\sim$\,2\,\textmu m thin membrane.\\
Dielectric coatings were applied in circular apertures with 2\,mm diameters centered on both outer SiC-facets. On one side the coating was designed for high reflectivity at the laser wavelength and high transmission for the 808\,nm pump wavelength. The other side was designed as an output coupling mirror with 97$\,\%$ reflectivity at the laser wavelength. The residual areas around the apertures were gold-plated as can be seen in Fig.\,\ref{fig:chip}\,(a.).
The lateral gain chip dimensions were 5\,\texttimes\,5\,mm$^{2}$ with a thickness of $\sim$\,1\,mm as depicted in the side perspective in Fig.\,\ref{fig:chip}\,(b.).

\subsection{Experimental setup}
\label{subsec:exp}
\begin{figure}
    \centering
    \includegraphics[width=1\linewidth]{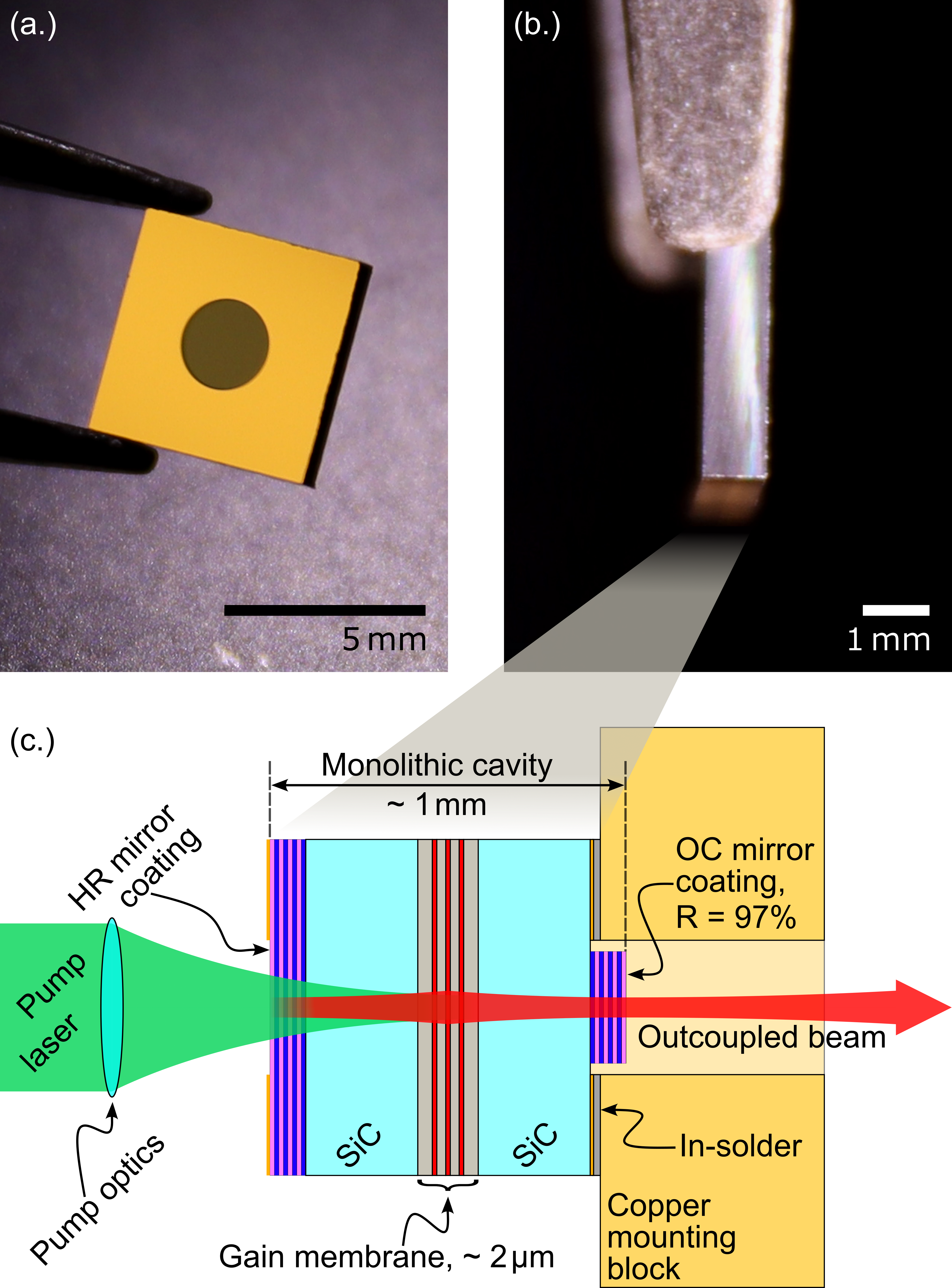}
    \caption{(a.) Photo of the microchip assembly with a size of 5\,\texttimes\,5\,\texttimes\,1\,mm$^{3}$ before soldering to the copper mounting block. (b.) Side perspective of the microchip assembly. (c.) Schematic of the pump arrangement and the microchip MECSEL with laser cavity (not to scale), soldered to the copper mount.}
    \label{fig:chip}
\end{figure}

This microchip assembly was soldered using indium to a gold-galvanized copper mounting block, which was then screwed to a thermo-electrically controlled sample holder. If not indicated otherwise, all measurements were performed at a fixed heat-sink temperature of $T_{hs}$\,=\,20°C. Figure\,\ref{fig:chip}\,(c.) illustrates a simplified schematic of the microchip MECSEL assembly. The laser is optically pumped with a fiber coupled BWT K808EA2FN 808\,nm diode laser. The fiber has a core diameter of 105\,\textmu m, a numerical aperture of 0.22, and it has a maximum output power of 10\,W. Due to the limited output power, thermal rollover was not observed in our investigations. A dual-lens pump-system was designed using a ray-tracing simulation software and consists of two achromatic lenses (Thorlabs AC254-045-B) with a focal length of 45\,mm.
The 4$\sigma$ diameter of the pump spot $\varnothing_\text{pump}$ was measured using a Gentec Electro-Optics Beamage-4M beam profiling camera to be $\varnothing_\text{pump}$\,=\,98.7\,\textmu m.
During all measurements $\varnothing_\text{pump}$ remained unchanged allowing for comparable results between the different measurements. As the microchip MECSELs outcoupling facet's coating was not designed to also reflect the 808\,nm pump light, the residual pump light, which is not absorbed in the semiconductor gain membrane is mainly transmitted and leaving the microchip. The weak internal back-reflections can be expected to contribute to additional minimal internal heating of the chip.
The laser chip itself consists of a plane-parallel cavity, which in principle is no stable resonator configuration. However, the formation of a thermal lens in the active region, which is a well-understood phenomenon in semiconductor lasers in general \cite{Lindberg.Strassner.ea_2005} and has been described in semiconductor microchip lasers \cite{Kemp.Maclean.ea_2006} as well as MECSELs \cite{Phung.Kahle.ea_2020} individually, stabilizes the laser.

\section{Laser characterization}
\label{sec:characterization}
\subsection{Output power}
\label{sec:output}

\begin{figure}[htbp]
    \centering
    \includegraphics[width=1\linewidth]{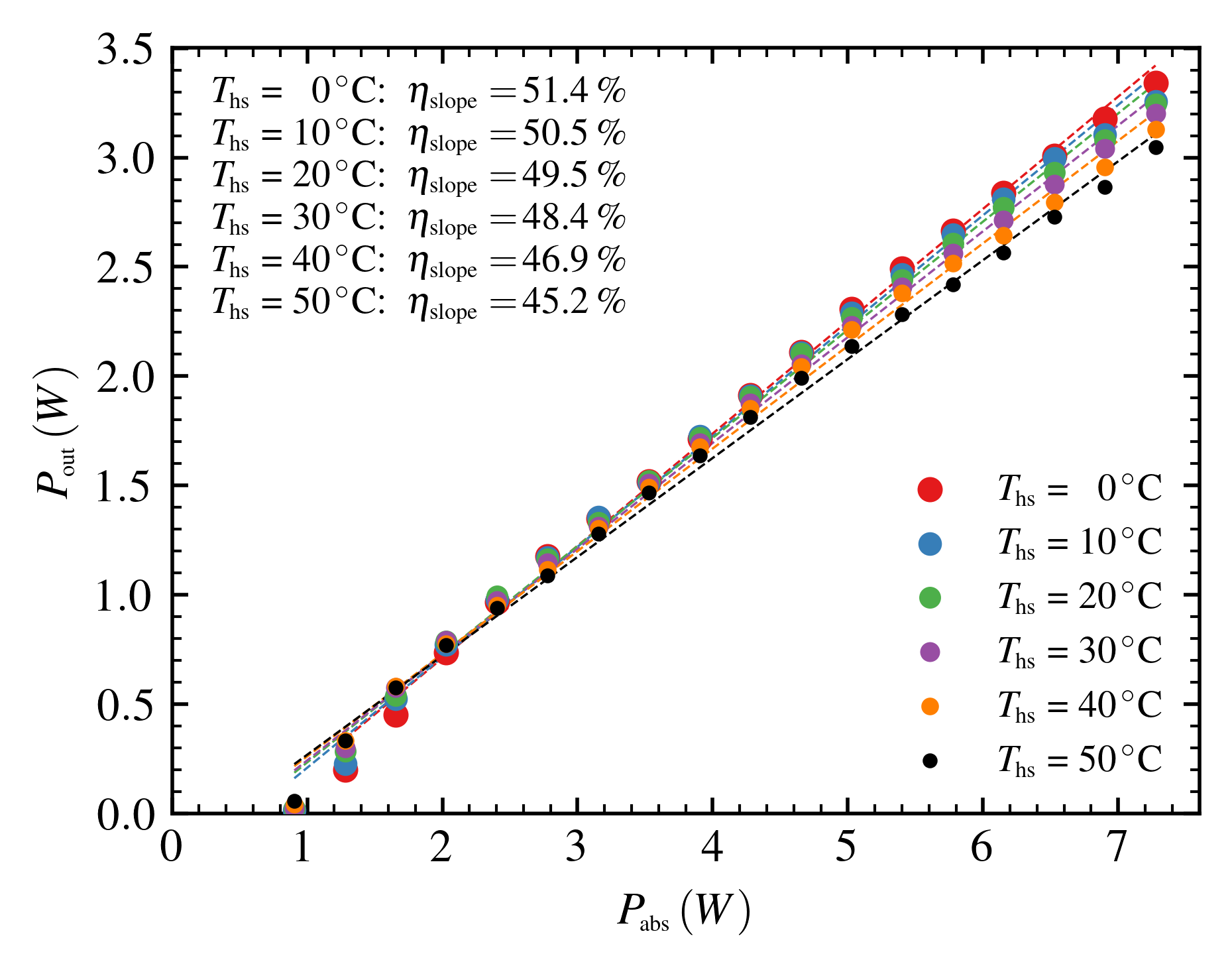}
    \caption{Output power of the microchip MECSEL (full circles) and linear fits (dashed lines) representing the slope efficiency $\eta_{\mathrm{slope}}$ versus absorbed pump power for different heat-sink temperatures.}
    \label{fig:powervspump}
\end{figure}

For the investigation of the laser chip's output characteristics, we measured the output power $P_{\mathrm{out}}$ versus the absorbed pump power $P_{\mathrm{abs}}$ for different heat-sink temperatures $T_{\mathrm{hs}}$, shown in Fig.\,\ref{fig:powervspump}.
For the measurement, the microchip MECSEL was pumped perpendicularly from the back side, a long-pass filter (Thorlabs FELH0850) in the output beam ensures that no pump power reaches the power meter (Newport 818P-30).\\ 
The laser threshold remained slightly below 1\,W of absorbed pump power, which did not change, even for increasing heat-sink temperatures. This indicates good charge carrier confinement and shows that even at $T_{\mathrm{hs}}$\,=\,50°C thermal escape does not reach a significant level to deteriorate the laser's threshold value. It is also a hint, that an evolving thermal lens needs to reach a certain strength, enabled by the excess absorbed pump power creating a stable cavity, which then allows laser emission.\\
The output power measurements were linearly fitted to obtain the slope efficiencies $\eta_{\mathrm{slope}}$, which are also presented in Fig.\,\ref{fig:powervspump}. It is important to mention that the output characteristics of VECSELs are linear above threshold and before the onset of thermal rollover effects. In contrast, our measurements show a deviation from this linear behavior due to the influence of the thermal lens, particularly at low powers. Nevertheless, we use the term ``slope efficiency'' here to enable comparison of our results with the literature.\\
For the lower and higher pump powers, the measured output powers are located slightly below the $\eta_{\mathrm{slope}}$ linear fits, while in the center range in between they are above. The thermal lens strongly depends on the temperature gradient between the hottest and coldest region in the chip. The hottest region is the pump spot center in the membrane. The coldest regions are the outer facets of the SiC heat-spreaders and can be assumed to have approximately heat-sink temperature $T_{hs}$. With increasing pump power, this temperature gradient increases resulting in a stronger thermal lens.\\
A stronger thermal lens results in a smaller beam diameter in the membrane, gradually improving the overlap with the pump spot, which leads to the effect of increasing efficiency with increasing $P_{\mathrm{abs}}$ in the first few data points in Fig.\,\ref{fig:powervspump}. For the highest absorbed pump powers in Fig.\,\ref{fig:powervspump} the measured powers are below the linear fits, as higher order modes start to oscillate and show a decreasing overlap with the pump spot in the active region compared to the TEM$_{00}$ modes.\\
As can be seen in Fig.\,\ref{fig:powervspump} the measurement at the highest $T_{hs}$ value shows the best performance at low pump powers and inferior values at higher pump powers. We attribute this behavior to the fact that, the evolving thermal lens enables better pump versus laser mode overlap at an earlier stage. Further, at higher pump powers the same effect leads to a accelerated decreasing of the mode diameter which results in an earlier worsening of pump versus laser mode overlap. Also, an interplay of the described effect with thermal escape of charge carriers could explain the lower performance at higher $T_{hs}$ values.
However the records for the differential efficiencies versus absorbed power are $67\,\%$ for in-well pumped VECSELs \cite{Beyertt.Brauch.ea_2007} and $59\,\%$ for barrier pumped VECSELs \cite{Keller.Sirbu.ea_2015}, the slope efficiency of $51.4\,\%$ reached in this work sets a new efficiency record for MECSELs.

\subsection{Emission wavelength of the laser}
\begin{figure}[htbp]
    \centering
    \includegraphics[width=1\linewidth]{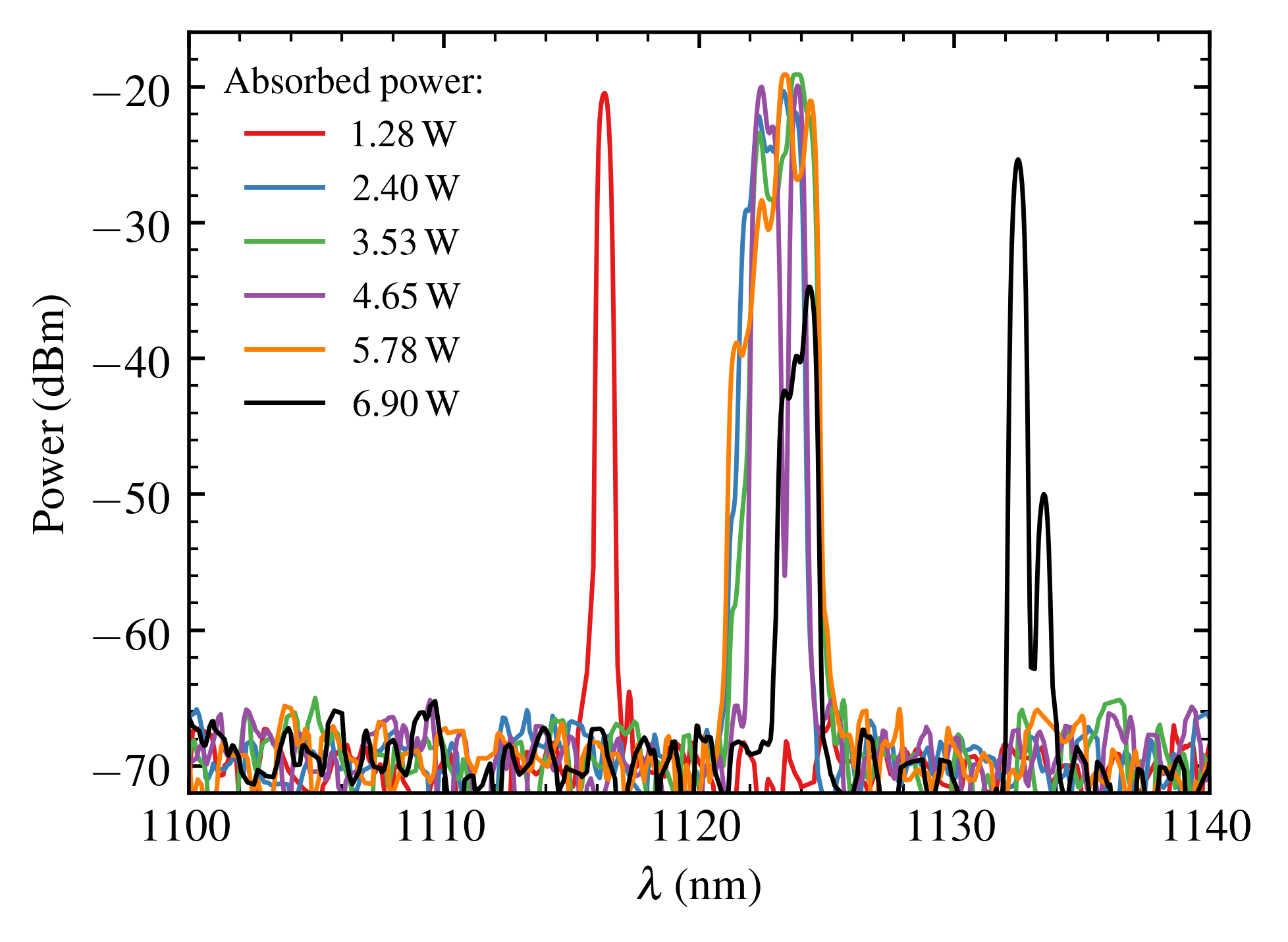}
    \caption{Emission spectra of the microchip MECSEL for different absorbed pump powers at 20$^\circ$C heat-sink temperature.}
    \label{fig:spectra}
\end{figure}

Figure\,\ref{fig:spectra} shows the output spectra of the microchip MECSEL at six different absorbed pump power values recorded with an Ando Q-6315A optical spectrum analyzer (OSA) with a resolution limit of 0.05\,nm. The laser was coupled into the OSA using a 100\,\textmu m fiber, which was placed directly in the divergent output beam without any fiber-coupling optics in order to avoid saturating the OSA. It should be noted that the power scale in Fig.\,\ref{fig:spectra} therefore does not correspond to the output power of the laser. 
It can be seen that the laser operates only in certain ranges of the spectrum while other wavelengths seem to be forbidden. This is most likely due to the use of SiC heat spreaders originating from different wafers. The nominal thickness of the double side polished SiC wafers is (500\,$\pm$\,25)\,\textmu m. The thickness variation within a single wafer is typically below 100\,nm and if neighboring pieces are used, this thickness variation is negligible. From wafer to wafer though, the thickness can strongly vary as the given tolerance of $\pm$\,25\,\textmu m indicates. \\
This explains, why we do not see an equidistant line pattern in the emission spectrum, as observed in free-running traditional MECSEL configurations such as in Ref.\,\cite{Kahle.Penttinen.ea_2019}. In the monolithic cavity of the microchip MECSEL with heat spreaders almost identical in thickness, this line pattern should be even stronger pronounced due to the enhanced reflectivity on the outer SiC facets compared to uncoated or even anti-reflection coated ones in an open external cavity. In turn, here lasing occurs where a beat node between the individual Fabry-Pérot transmission of each heat spreader creates the least interference losses.\\
We measured the thicknesses of two SiC pieces, originating from the same two wafers used as the transparent heat-spreaders of the microchip MECSEL, to be 502.6\,\textmu m and 483.5\,\textmu m. With these values the transmission spectrum of two beating etalons was simulated. The resulting spectrum has a free-spectral range (FSR) of $\sim$\,0.5\,nm, which represent the Fabry-Pérot resonances of the individual heat-spreaders, and a beating of $\sim$\,14\,nm between the two heat-spreaders. In Fig.\,\ref{fig:spectra}, the free spectral range as well as beating between the two heat-spreaders can be observed. Even though the values obtained in the simulation don't precisely match the distances between the peaks in the measurement, the simulations give a good indication for the origin of the observed behavior. 

\subsection{Thermal resistance}
For the measurement of the microchip MECSEL's thermal resistance $R_{\mathrm{th}}$, the approach described in \cite{Heinen.Zhang.ea_2012} had to be modified, using photoluminescence (PL) instead of laser emission spectra due to the wavelength-fixing FSR and beating effects described in the previous section under lasing conditions. The thermal resistance can be determined by 
\begin{equation}
    R_{\mathrm{th}} = \frac{\nicefrac{\partial\lambda}{\partial P_{\mathrm{diss}}}}{\nicefrac{\partial\lambda}{\partial T_{\mathrm{hs}}}} = \frac{\partial T_{\mathrm{hs}}}{\partial P_{\mathrm{diss}}}~,
\end{equation}
where ${\partial\lambda}/{\partial P_{\mathrm{diss}}}$ is the wavelength shift with dissipated power and ${\partial\lambda}/{\partial T_{\mathrm{hs}}}$ is the wavelength shift per heat-sink temperature.\\
For the measurements determining the wavelength shifts, we used a chip from the same wafer as the microchip MECSEL, but with an anti-reflective coated outcoupling facet instead. The chip was soldered to a similar heat-sink as the microchip MECSEL with the same solder to get a similar thermal behavior. This chip was used without an external mirror to avoid lasing and only observe the photoluminescence.\\
For the wavelength shift versus heat-sink temperature, the chip was pumped with a power of 0.8\,W and a duty cycle of 10\,\% was applied to the pump beam with a chopper wheel. This ensures that the temperature in the membrane is fairly close to the heat-sink temperature in the unpumped and pumped configuration. PL spectra were recorded for six different temperatures between $10^\circ$C and $60^\circ$C using the same OSA as for the emission spectra measurements.\\
For the wavelength shift with dissipated power ${\partial\lambda}/{\partial P_{\mathrm{diss}}}$, the chopper wheel was removed and PL spectra were recorded while the chip was pumped with absorbed pump powers between $0.9\,$W and $3.5\,$W
at a fixed heat-sink temperature of $20^\circ$C. \\
The dissipated power is generally defined as the absorbed pump power reduced by the output power: $P_{\mathrm{diss}}$\,=\,$P_{\mathrm{abs}}$\,-\,$P_{\mathrm{out}}$. In our case, the dissipated power was neglected because the total output power contained in the photoluminescence is small.\\
The recorded PL spectra were fitted with asymmetric Lorentz oscillators. Lorentz oscillators were used for the fitting because they accurately describe resonant processes and energy dissipation, reflecting the physical mechanisms underlying spectral line shapes. The longer wavelength at full-width half-maximum (FWHM)  was extracted from the fits. \\
This value was used instead of the peak wavelength, because it represents a good standardized way to extract a wavelength and depicts an assumption of decent accuracy for the spectral location of the gain maximum, which is, where the laser would oscillate, if not influenced by spectral filters creating forbidden areas. But exactly this was the case, as can be also seen in Fig.\,\ref{fig:spectra}. Only at lowest and highest pump powers the emission wavelength of the microchip MECSEL shifts. During all other measurements it appears to be fixed.\\
For the wavelength shift per heat-sink temperature ${\partial\lambda}/{\partial T_{\mathrm{hs}}}$, the linear fit to the data reveals ${\partial\lambda}/{\partial T_{\mathrm{hs}}}$\,=\,0.388\,nm/K, the wavelength shift per dissipated power yields ${\partial\lambda}/{\partial P_{\mathrm{diss}}}$\,=\,0.595\,nm/W. This results in a thermal resistance of $R_{\mathrm{th}}$\,=\,1.534\,K/W, which is in the range of previously determined values for the thermal resistance in MECSELs \cite{Schuchter.Huwyler.ea_2024, Phung.TatarMathes.ea_2022}.

\subsection{Beam quality investigations}
In order to investigate the beam quality of the microchip MECSEL, a Gentec Electro-Optics Beamage-4M beam profiling camera (sensor size $11.264\,\times\,11.264$\,mm$^2$) was used. The camera was first placed directly inside the output beam with a long-pass filter and an ND-filter to reduce the power reaching the camera chip. 

\subsubsection{Beam profiles}

\begin{figure}[htbp]
    \centering
    \includegraphics[width=1\linewidth]{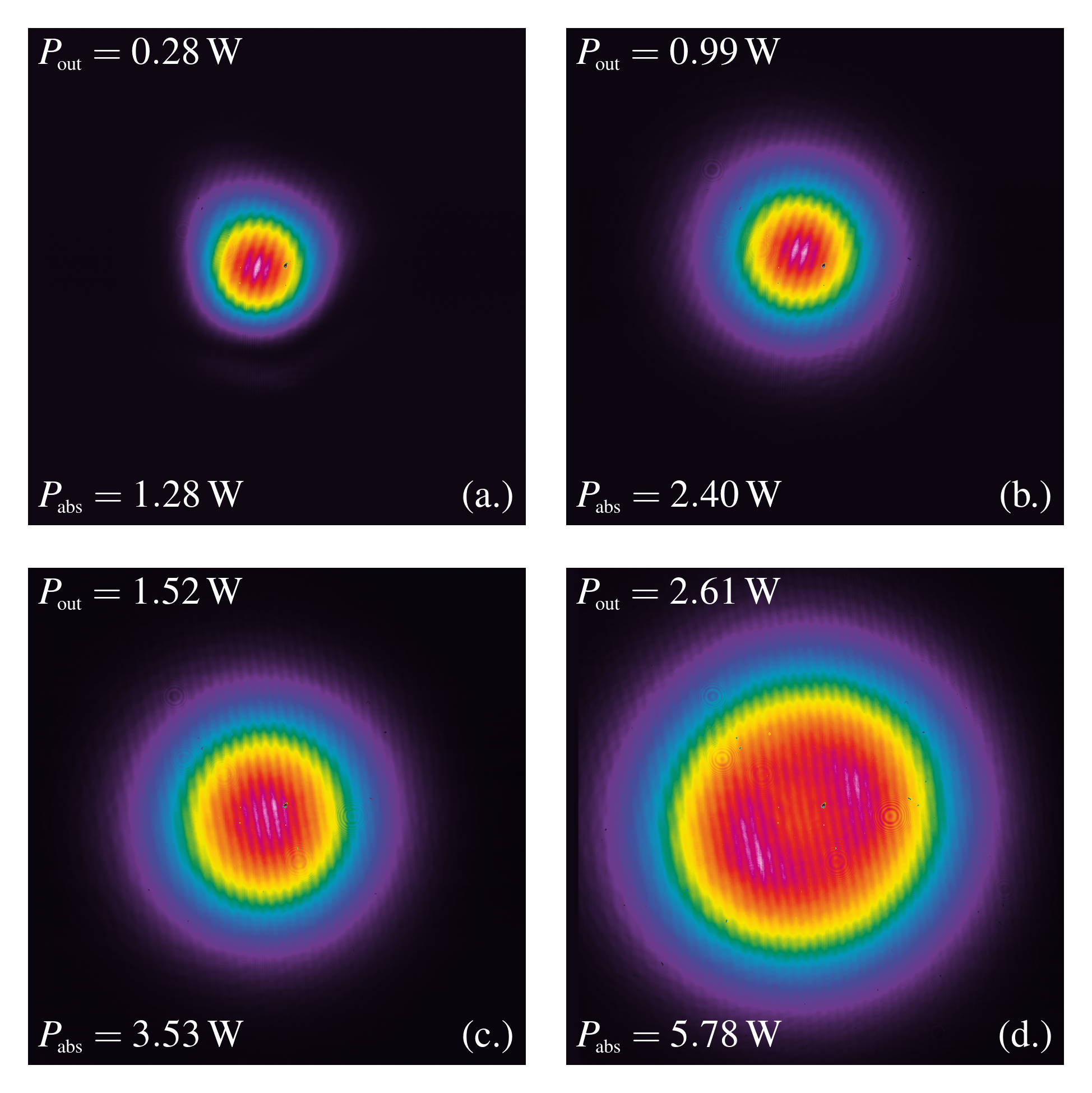}
    \caption{Beam profiles recorded at different absorbed pump powers without changing the position of the beam profile camera.}
    \label{fig:beams}
\end{figure}

Figure\,\ref{fig:beams} shows the beam profile of the laser emission for different absorbed pump powers.
In Fig.\,\ref{fig:beams}\,(a.), the laser is still close to its threshold and the thermal lens is weak, leading to a non-perfect circular beam. In Fig.\,\ref{fig:beams}\,(b.) and (c.), the beam profile looks almost perfectly circular. In Fig.\,\ref{fig:beams}\,(d.), it can be seen that higher order transverse modes start to overlap with the fundamental TEM$_{00}$ mode. The distinct stripe-pattern in the beam profiles originates from interference between optical filters in the beam path.

\subsubsection{Beam quality factor}
\begin{figure}[htbp]
    \centering
    \includegraphics[width=1\linewidth]{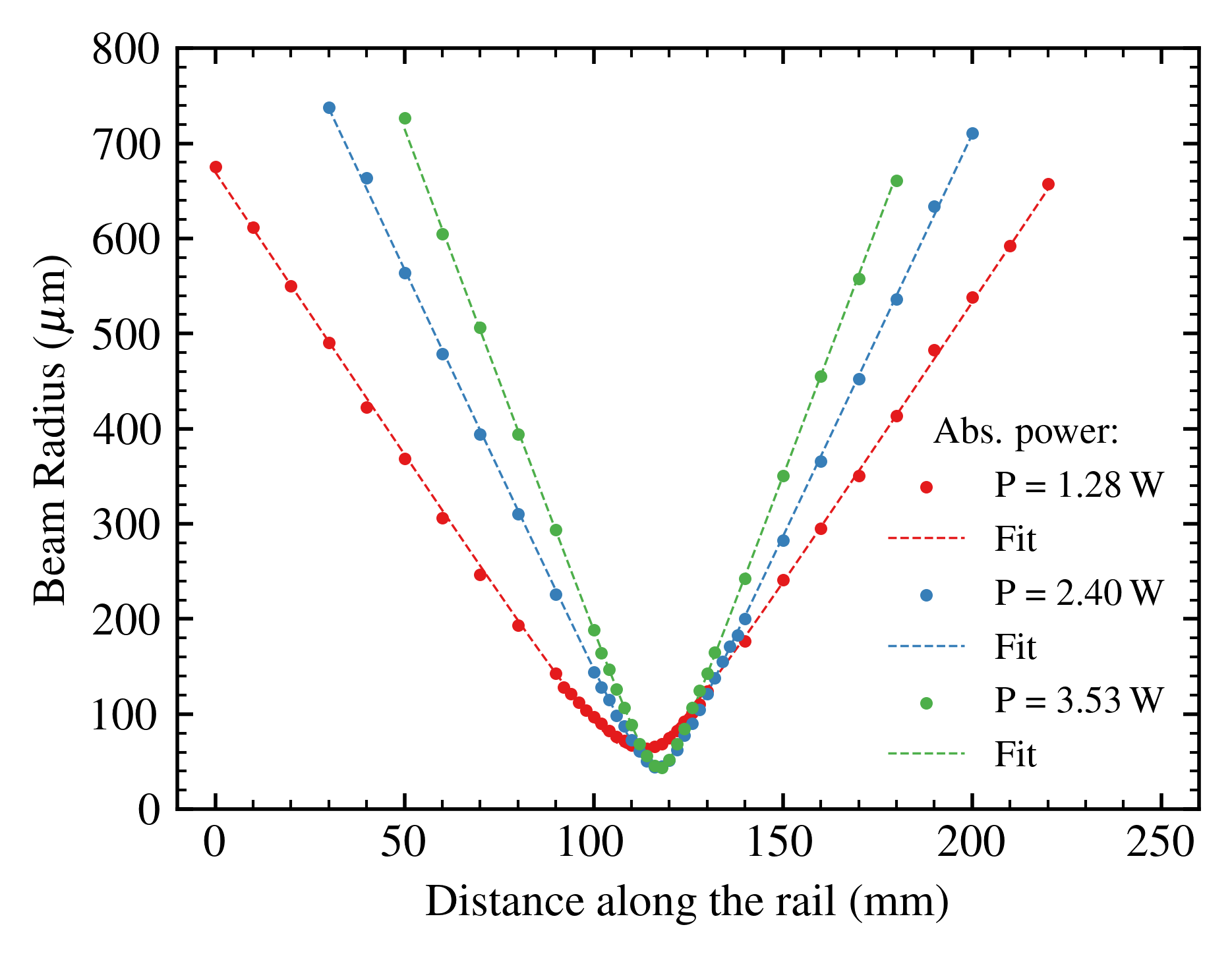}
    \caption{Output beam radius along the measurement rail (marks) and fits for determining $M^2$ (dashed lines) at three different absorbed pump powers. The three measurements correspond to the first three images in Fig.\,\ref{fig:beams}.}
    \label{fig:m2}
\end{figure}

For the determination of the beam quality factor $M^2$, the laser's output beam was collimated with an $f$\,=\,300\,mm lens, weakened with a 95\,\%/5\,\% beam-splitter and then focused again with an $f$\,=\,250\,mm lens. The beam profiling camera was mounted on a rail to measure the second-moment width ($D4\sigma$) of the beam along the beam path. Around the focus, the density of data points was increased collecting enough measurements within the Rayleigh-length for valid fit results. The so determined beam radius was then fitted to 
\begin{equation}
    w^2(z) = w_0^2 + \left(M^2\right)^2 \left(\frac{\lambda}{\pi\cdot w_0}\right)^2 (z-z_o)^2~,
\label{eq:m2fiteq}
\end{equation}
where $w^2(z)$ is the beam radius at a given distance $z$ along the rail. The beam waist $w_0$, the beam quality factor $M^2$, and the location of the beam waist $z_0$ were used as fit parameters. Figure\,\ref{fig:m2} shows the measurements and fit curves for three different pump powers. 
It can be seen that we get a good match between the measurements and fits. For increasing pump powers, the divergence angle of the beam increases while the beam waist decreases. The slight shift to the right can be explained with an imperfect collimation of the beam by the first lens, however this does not influence the results.

\subsubsection{Beam divergence angle}
Besides $M^2$, the beam divergence angle $\theta_{\text{D}}$ was measured by removing the lenses from the beam path and measuring the diameter of the lasers output at 13 different locations along the rail. $\theta_{\text{D}}$ can then be obtained from the slope of a linear fit of the measured diameters. Figure\,\ref{fig:multiplot} shows the measured results for $M^2$ and $\theta_\text{D}$. From those two values, the beam waist in the chip $w_{0,\,\mathrm{microchip}}$ can be calculated by 
\begin{equation}
    w_{0,\,\mathrm{chip}} = \frac{M^2\cdot\lambda}{\pi\cdot\theta_{\text{D}}}.
\end{equation}
The results are shown in Fig.\,\ref{fig:multiplot} as well.

\subsubsection{Thermal lens}
\begin{figure}[htbp]
    \centering
    \includegraphics[width=1\linewidth]{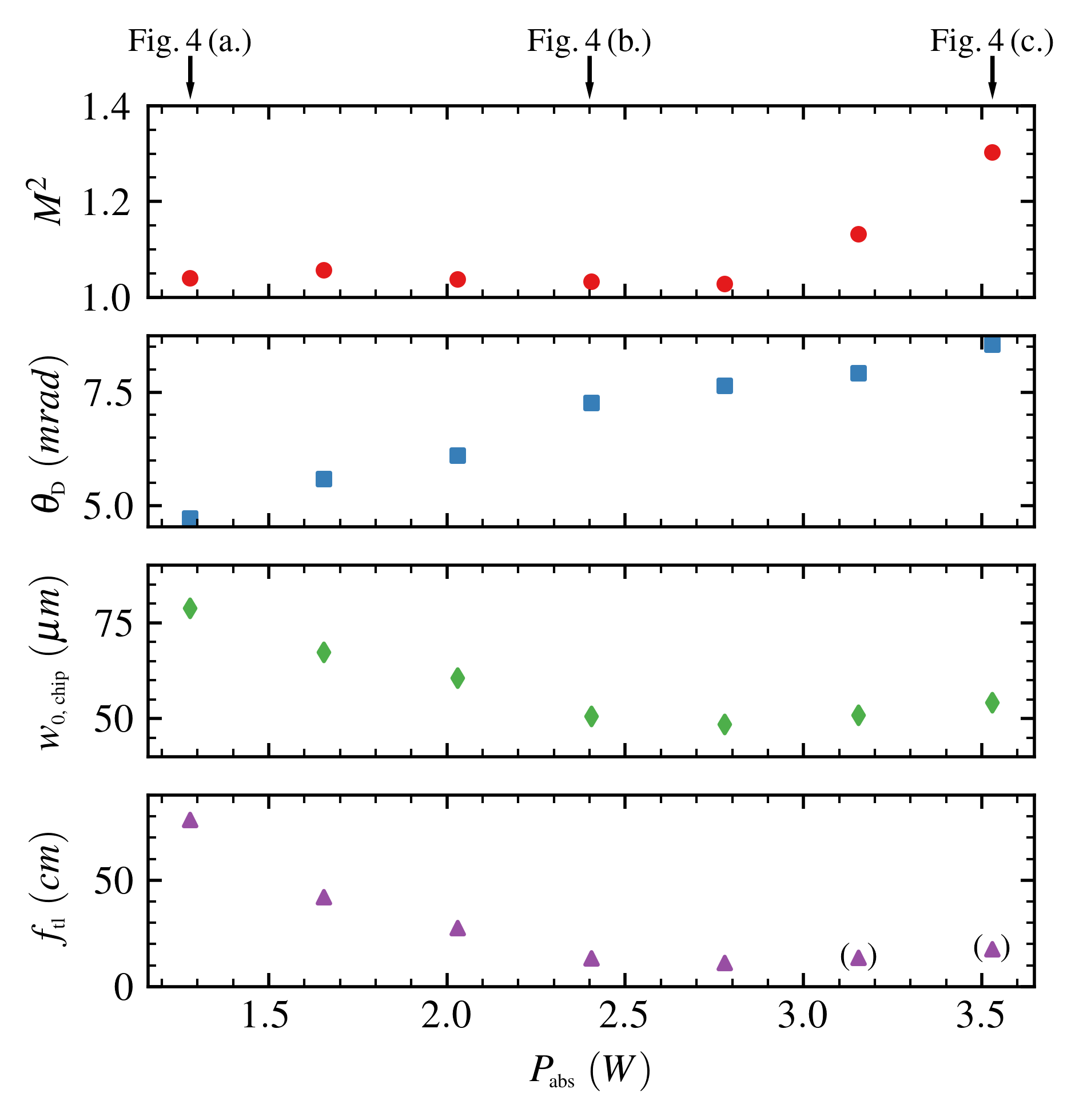}
    \caption{Beam quality factor $M^2$, divergence angle $\theta_D$, calculated beam waist in the laser chip $w_{0,~\text{chip}}$ and calculated focal length of the thermals lens $f_{tl}$ versus absorbed pump power $P_\text{abs}$. Because the last two values for the focal length of the thermals lens $f_mathrm{tl}$ are not reliable, they were put in brackets. It is shown which data points correspond to the beam profiles in Fig.\,\ref{fig:beams}.}
    \label{fig:multiplot}
\end{figure}

Further, we used the transfer matrix method to estimate the focal length of the thermal lens forming inside the laser chip from the calculated beam waists at different powers. The results are included in Fig.\,\ref{fig:multiplot} as well.
Further, it is pointed out which data points correspond to the beam profiles in Fig.\,\ref{fig:beams}. For low powers the thermal lens with a focal length of $f_{\mathrm{tl}}$\,=\,78.52\,cm is relatively weak. $f_{\mathrm{tl}}$ shortens with increasing absorbed pump power until it reaches its minimum of $f_{\mathrm{tl}}$\,=\,11.29\,cm at 3.15\,W. For further increasing powers, the values are not reliable because the transfer matrix is defined for Gaussian beams only and the increasing value of $M^2$ indicates the beginning presence of higher-order modes. This is why the last two values for the focal length of the thermal lens in Fig.\,\ref{fig:multiplot} were put in brackets. Nevertheless, as the divergence angle $\theta_{\text{D}}$ increases further one can assume the thermal lens to strengthen as well, while it also seems to run into saturation. Such a behavior was also observed in previous work \cite{Phung.Kahle.ea_2020}.

\subsubsection{Polarization}
\begin{figure}[htbp]
    \centering
    \includegraphics[width=1\linewidth]{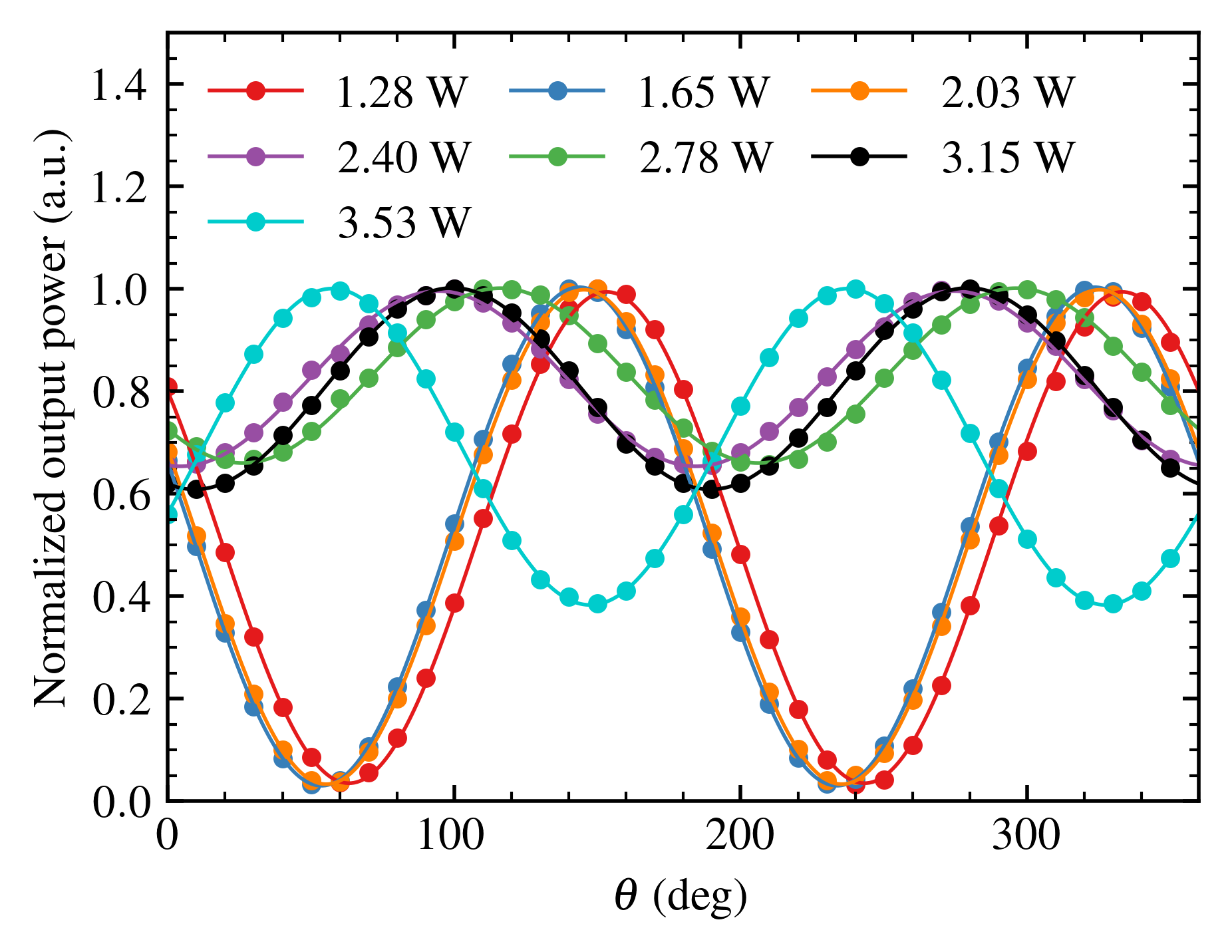}
    \caption{Normalized output power versus rotation angle of the polarization filter (marks) and fits (lines) for different absorbed pump powers.}
    \label{fig:polarization}
\end{figure}

Another important parameter which allows to assess the quality of a laser is its polarization. A linear polarizer (LPNIRE100-B from Thorlabs) was used to investigate the outcoupled beam. The polarizer was rotated in 10° steps by 360° in total and the transmitted laser power was measured for different absorbed pump powers. The measured output power versus rotation measurements were fitted using 
\begin{equation}
    P(\theta) = P_0 + P_1 \cos^2(\theta - \theta_0)~,
\end{equation}
where $P_0$ is the unpolarized fraction of the output power, $P_1$ is the amplitude of the polarized fraction of the output power and $\theta_0$ is an angular offset. 
A well known parameter to assess the quality of polarization is the degree of polarization ($DOP$) \cite{AlQasimi.Korotkova.ea_2007, Kikuchi_2001}, which is calculated as 
\begin{equation}
    DOP = \frac{P_{\mathrm{max}}-P_{\mathrm{min}}}{P_{\mathrm{max}}+P_{\mathrm{min}}} = \frac{P_1}{P_1+2P_0}~.
    \label{eq:dop}
\end{equation}
Figure\,\ref{fig:polarization} shows an exemplary measurement of the polarization for different absorbed pump powers and the fits.
The results for the DOP are presented in Tab.\,\ref{tab:pol}. It can be seen that the DOP is close to unity for lower absorbed pump powers but decreases for higher powers. This could result from two differently polarized modes with different thresholds.
\begin{table}[htbp]
    \centering
    \caption{Degree of polarization of the microchip MECSEL for different absorbed pump powers}
    \begin{tabular}{ccc}
        \hline
        Absorbed pump power & DOP &  $\theta_0$ \\
        \hline
        1.28\,W & 93.14\,\% & 153.25° \\
        1.65\,W & 94.11\,\% & 143.65° \\
        2.03\,W & 93.53\,\% & 145.31° \\
        2.40\,W & 20.74\,\% & 94.43° \\
        2.78\,W & 20.54\,\% & 116.10° \\
        3.15\,W & 24.41\,\% & 99.42° \\
        3.53\,W & 44.62\,\% & 57.42° \\
        \hline
    \end{tabular}
    \label{tab:pol}
    %$^\textit{a}$Only quadratic line elements are included here.
\end{table}

\section{Conclusion}
\subsection{Summary}
A microchip MECSEL was presented for the first time to our knowledge. An InGaAsP gain membrane is sandwiched between two silicon carbide heat spreaders with one highly reflective coating on one and an outcoupling coating on the other facet. Similar to VCSELs, the resonator stabilizes itself by the formation of a thermal lens. The microchip MECSEL operates at around 1123\,nm and shows an output power of 3.2\,W at 7.28\,W absorbed pump power with a heat-sink temperature of $T_{\mathrm{hs}}$\,=\,20°C. Slope efficiencies of more then 51\,\% were observed. We examined the emission wavelength of the microchip MECSEL. A distinct beating between the two Fabry-Pérot etalon resonances of the two heat-spreaders creates spectrally forbidden regions. Our investigations on the thermal resistance showed a good agreement with previously reported values. We determined the beam-quality factor $M^2$ of the laser at different absorbed pump powers, showing single transverse mode operation at an output power of more than 1\,W. From the $M^2$ and the beam divergence angle, we determined the beam waist in the gain membrane and used the transfer matrix method to estimate the focal length of the thermal lens forming in the chip at different absorbed pump powers. The focal length of the thermal lens decreases from 79\,cm to 11\,cm with increasing absorbed pump power. It was further shown, that the microchip MECSEL is well polarized for low absorbed pump powers, however the degree of polarization decreases when a differently polarized mode starts to oscillate in the cavity, even though both modes are TEM$_{00}$ because $M^2$ is still close to one.

\subsection{Outlook and discussion}
The concept of microchip MECSELs presented in this work could, in principal, be applied to all known material systems used for semiconductor lasers nowadays to access a wide wavelength range. The manufacturing process can be adopted to wafer-scale \cite{Cole.Nguyen.ea_2022} making microchip MECSELs an interesting approach for industrial production for commercial products.\\
Integration of second harmonic crystals into the solid-state cavity is also possible \cite{Sinclair_1999} and could as well be performed in wafer-scale or as single chip. It will enable extremely compact solutions for wavelengths yet uncovered with highest beam quality lasers in the watt level.\\
A resonant grating mirror \cite{Bisson.Parriaux.ea_2006}, also similar to the high contrast gratings described in Ref.\,\cite{Iakovlev.Walczak.ea_2014}, could be used to stabilize and fix the polarization, overwriting the internal mechanisms leading to a material-preferred polarization orientation other than one favored by a laser manufacturer.\\
The use of micro-lensed heat spreaders was already shown \cite{Park.Jeon_2006,Laurand.Lee.ea_2007,Laurand.Lee.ea_2009}. A microchip MECSEL could benefit from such an approach. A micro-lens, added afterwards by optical bonding or shaped into one or both of the heat spreaders, would provide a stable cavity configuration from the beginning. With a hemispherical or confocal-concentric configuration, the laser would not have to "wait" for a thermal lens to stabilize the plane-parallel cavity. This would likely lower the laser threshold because the losses would be lower compared to a plane-parallel configuration. A hemispherical/confocal-concentric cavity configuration would possibly also prevent the early appearance of higher transverse modes and shift this behavior closer to the thermal rollover of the laser, where it typically occurs.\\
These advancements highlight the potential of microchip MECSELs for filling the gap between commercially available VCSELs, which cover the needs for very small (single device $\sim$\,10$^{-3}$\,mm$^3$), low power, and good beam quality, (e.g. Ref.\,\cite{Liu.Wang.ea_2024}), or array applications (e.g. Ref.\,\cite{Ott.Stange.ea_2025}), and classical VECSELs \cite{Guina.Rantamaeki.ea_2017,Jetter.Michler_2021}, representing an extremely versatile platform providing high power, tunability by design and potentially intra-cavity wavelength multiplying and polarization stabilization, while maintaining excellent beam quality.

\section*{Acknowledgments}
The authors thank Prof.\,Dr.~Peter~Unger for the provided support and access to the lab at Ulm University, where the experiments presented in this work were carried out. Furthermore, the authors want to thank Alina Tonn from the Jagsttal-Gymnasium Möckmühl for assisting with the measurements during her high-school internship.

\subsubsection*{Disclosures}
The authors declare no conflict of interest.
%\newpage
% Bibliography
\bibliography{MonoMECSEL}

\bibliographyfullrefs{MonoMECSEL}

\ifthenelse{\equal{\journalref}{aop}}{%
\section*{Author Biographies}
\begingroup
\setlength\intextsep{0pt}
\begin{minipage}[t][6.3cm][t]{1.0\textwidth} % Adjust height [6.3cm] as required for separation of bio photos.
  \begin{wrapfigure}{L}{0.25\textwidth}
    \includegraphics[width=0.25\textwidth]{john_smith.eps}
  \end{wrapfigure}
  \noindent
  {\bfseries John Smith} received his BSc (Mathematics) in 2000 from The University of Maryland. His research interests include lasers and optics.
\end{minipage}
\begin{minipage}{1.0\textwidth}
  \begin{wrapfigure}{L}{0.25\textwidth}
    \includegraphics[width=0.25\textwidth]{alice_smith.eps}
  \end{wrapfigure}
  \noindent
  {\bfseries Alice Smith} also received her BSc (Mathematics) in 2000 from The University of Maryland. Her research interests also include lasers and optics.
\end{minipage}
\endgroup
}{}

\end{document}